\documentclass[conference]{IEEEtran}
\IEEEoverridecommandlockouts
\usepackage{geometry}
\usepackage{cite}
\usepackage{amsmath,amssymb,amsfonts}
\usepackage{algorithmic}
\usepackage{graphicx}
\usepackage{textcomp}
\usepackage{xcolor}
\def\BibTeX{{\rm B\kern-.05em{\sc i\kern-.025em b}\kern-.08em
    T\kern-.1667em\lower.7ex\hbox{E}\kern-.125emX}}

\usepackage{mathtools}

\usepackage{steinmetz}
\usepackage{authblk}
\newcommand{\minus}{\scalebox{0.75}[1.0]{$-$}}
\usepackage[ruled,norelsize]{algorithm2e}
\makeatletter
\newcommand{\removelatexerror}{\let\@latex@error\@gobble}
\makeatother

\usepackage{blindtext}

\usepackage{caption}
\usepackage{subcaption}

\begin{document}

\title{Rate Splitting Multiple Access for Multi-Antenna Multi-Carrier Joint Communications and Jamming\\
	\thanks{This work was supported by the Engineering and Physical Sciences Research Council of the UK (EPSRC) Grant number EP/S026657/1, and the UK MOD University Defence Research Collaboration (UDRC) in Signal Processing.}}
\author{
	Onur~Dizdar and~Bruno~Clerckx\\
	Department of Electrical and Electronic Engineering, Imperial College London\\
	Email: \{o.dizdar, b.clerckx\}@imperial.ac.uk}

\maketitle
\begin{abstract}
	Rate-Splitting Multiple Access (RSMA) is a robust multiple access scheme for downlink multi-antenna wireless networks. 
	In this work, we investigate a novel application of RSMA for joint communications and jamming with a Multi-Carrier (MC) waveform in Multiple Input Single Output (MISO) Broadcast Channel (BC). Our aim is to simultaneously communicate with Information Users (IU) and jam Adversarial Users (AU) to disrupt their communications in a setting where all users perform broadband communications by MC waveforms in their respective networks.
	We consider the practical setting of imperfect CSI at transmitter (CSIT) for the IUs and statistical CSIT for AUs.
	The optimal information and jamming precoders are designed to maximize the sum-rate under jamming power constraints on the pilot subcarriers of AUs, a jamming method considered to be among the most destructive methods for MC waveforms under the considered system model. 
	We compare the sum-rate performance of RSMA and Space Division Multiple Access (SDMA) schemes by numerical results to demonstrate that RSMA achieves a significant sum-rate gain compared to SDMA.  
\end{abstract}

\section{Introduction}
Rate-Splitting Multiple Access (RSMA) is a multiple access scheme based on the concept of Rate-Splitting (RS) and linear precoding for multi-antenna multi-user communications. RSMA splits user messages into common and private parts, and encodes the common parts into one or several common streams while encoding the private parts into separate streams. The streams are precoded using the available (perfect or imperfect) Channel State Information at the Transmitter (CSIT), superposed and transmitted via the Multi-Input Multi-Output (MIMO) or Multi-Input Single-Output (MISO) channel \cite{clerckx_2016}. All the receivers then decode the common stream(s), perform Successive Interference Cancellation (SIC) and then decode their respective private streams. Each receiver reconstructs its original message from the part of its message embedded in the common stream(s) and its intended private stream.

RSMA manages multi-user interference by allowing the interference to be partially decoded and partially treated as noise. RSMA has been shown to embrace and outperform existing multiple access schemes, i.e., Space Division Multiple Access (SDMA), Non-Orthogonal Multiple Access (NOMA), Orthogonal Multiple Access (OMA) and multicasting. The sum-rate performance of
RSMA has been demonstrated to be robust and to surpass the performance of SDMA and
NOMA under perfect and imperfect CSIT in numerous works 
\cite{clerckx_2016,clerckx_2019,Joudeh_2016,mao_2018,mao_2019_2}.

With increasing number of systems and applications, the Radio-Frequency (RF) spectrum has become a congested and contested environment. Both commercial and military systems require broadband communications to meet the data requirements for the advancing applications. In such a congested and contested environment, efficient use and sharing of spectrum is of high importance, especially for military communications with strict reliability and robustness requirements.    
An equally critical target in military communications is denying the Adversarial Users (AUs) of service. In this work, we are interested in jamming methods targeting Multi-Carrier (MC) waveforms used for broadband communications. Among the jamming methods for MC waveforms, pilot subcarrier jamming is accepted to be one of the most destructive ones which consist of Artificial Noise (AN) ({\sl i.e.,} excluding the methods which use valid data signals to cause misconception in the network or attack the synchronisation of the waveform). Pilot jamming aims to disrupt the channel estimation procedure of the affected user to prevent error-free detection and decoding of its intended messages \cite{Shahriar_2015, clancy_2011, miller_2012, patel_2004, han_2008}.

In this work, we consider a scenario where a multi-antenna military system aims to communicate in an RF-congested environment while simultaneously performing jamming to AUs in the same spectrum. We use RSMA for multi-carrier multiple-access communications. Our aim is to identify the performance benefits of RSMA for joint communications and jamming and obtain optimal precoders to maximize the mutual information for communications with Information Users (IUs) while performing jamming on pilot subcarriers of AUs efficiently.
We consider the practical and realistic scenario of imperfect Channel State Information at Transmitter (CSIT) for the IUs and statistical CSIT for AUs, since obtaining an accurate channel estimate for the AUs is generally not feasible \cite{karlsson_2017}. We give a formulation for MC waveforms to solve the abovementioned problem, which turns out to be non-convex. 
We propose a Successive Convex Approximation (SCA) based algorithm, similar to the algorithm for single carrier waveforms in \cite{Mao_2016}, combined with the Sample Average Approximated Weighted Minimum Mean Square (SAA-WMMSE) algorithm in \cite{Joudeh_2016} to solve the problem. 
We demonstrate by simulations that RSMA achieves a significant sum-rate performance over SDMA. 

\textit{Notations:} Vectors are denoted by bold lowercase letters and matrices are denoted by bold uppercase letters. 
The operations $|.|$ and $||.||$ denote the absolute value of a scalar and $l_{2}$-norm of a vector, respectively, unless stated otherwise. 
$\mathbf{a}^{H}$ denotes the Hermitian transpose of a vector $\mathbf{a}$. $\mathcal{CN}(0,\sigma^{2})$ denotes the Circularly Symmetric Complex Gaussian distribution with zero mean and variance $\sigma^{2}$. $\mathbf{I}_{n}$ denotes the $n$-by-$n$ identity matrix. The operator $\mathrm{tr}(.)$ denotes the trace operation.
The operator $\mathrm{Diag}(\mathrm{X}_{1}, \ldots, \mathrm{X}_{K})$ builds a matrix $\mathrm{X}$ by placing the matrices $\mathrm{X}_{1}$, $\ldots$, $\mathrm{X}_{K}$ diagonally and setting all other elements to zero. 

\section{System Model}
\label{sec:system}
\begin{figure}[t!]	
	\centerline{\includegraphics[width=2.5in,height=2.5in,keepaspectratio]{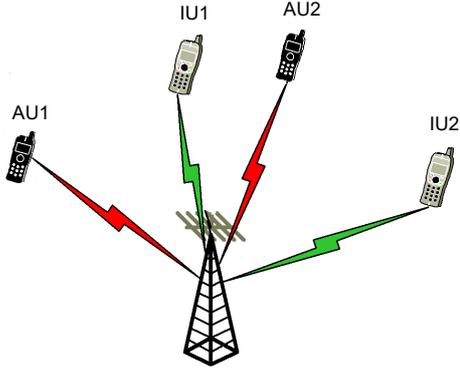}}
	\vspace{-0.2cm}
	\caption{System model.}
	\label{fig:system}
	\vspace{-0.8cm}
\end{figure}

We consider a Multi-Input Single-Output (MISO) Broadcast Channel (BC) setting consisting of one transmitter with $n_{t}$ transmit antennas, which aims to perform communications and jamming simultaneously in an RF-congested environment, as illustrated in Fig.~\ref{fig:system}. The transmitter serves $K$ single-antenna IUs indexed by $\mathcal{K}=\left\lbrace 1,2,\ldots,K\right\rbrace$, while simultaneously performing jamming on $L$ single-antenna AUs indexed by $\mathcal{L}=\left\lbrace 1,2,\ldots,L\right\rbrace $. 
We assume that the IUs in the network are operating in the same frequency band as the AUs. The transmitter employs an MC waveform to communicate with the IUs, while the AUs also use an MC waveform to communicate in their corresponding separate networks. We define the set of subcarrier indexes in the signal band as $\mathcal{S}=\left\lbrace 1,2,\ldots,N\right\rbrace $. Also, we define the set of pilot subcarriers of AUs in the same signal band as $\mathcal{S}_{p} \subset \mathcal{S}$.  

We employ 1-layer RSMA \cite{mao_2018} to perform multiple-access communications in the considered setting.
RSMA relies on splitting the user messages at the transmitter side. The messages intended for the users, $W_{k,n}$, are split into common and private parts, i.e., $W_{c,k,n}$ and $W_{p,k,n}$, $\forall k\in \mathcal{K}$, $n\in \mathcal{S}$. The common parts of the messages of all users are combined into the common message $W_{c,n}$. 
The common message $W_{c,n}$ and the private messages are independently encoded into streams $s_{c,n}$ and $s_{k,n}$, respectively. 
Jamming is performed on subcarrier-$n$ of AU-$l$ using the AN signal $s^{e}_{l,n}$, $\forall l \in \mathcal{L}$ and $\forall n \in \mathcal{S}_{p}$. 
We assume that each subcarrier is assigned a separate precoder. The MC transmit signal for RSMA is 
\vspace{-0.3cm}
\begin{align}
	x_{n}=\mathbf{p}_{c,n}s^{f}_{c,n}+\sum_{k=1}^{K}\mathbf{p}_{k,n}s^{f}_{k,n}+\sum_{l=1}^{L}\mathbf{f}_{l,n}s^{e}_{l,n}, \   n\in \mathcal{S}.  \nonumber
	\vspace{-0.3cm}
\end{align}
The vectors $\mathbf{p}_{c,n} \in\mathbb{C}^{n_t}$ and $\mathbf{p}_{k,n} \in\mathbb{C}^{n_t}$ are the linear precoders applied to the common stream and the private stream of IU-$k$ on subcarrier $n$, $\forall k \in \mathcal{K}$ and $\forall n \in \mathcal{S}$. 
The precoder $\mathbf{f}_{l,n}$ is used to transmit AN to AU-$l$, $\forall l \in \mathcal{L}$. The communications signals $s^{f}_{c,n}$ and $s^{f}_{k,n}$ and jamming signals $s^{e}_{l,n}$ are chosen independently from a Gaussian alphabet for theoretical analysis. We also assume that the streams have unit power, so that \mbox{$\mathbb{E}\left\lbrace \mathbf{s}_{n}(\mathbf{s})^{H}\right\rbrace =\mathbf{I}_{K+L+1}$}, where \mbox{$\mathbf{s}_{n}=[s^{f}_{c,n}, s^{f}_{1,n}, \ldots, s^{f}_{K,n}, s^{e}_{1,n}, \ldots, s^{e}_{L,n}]$}. An average transmit power constraint is set as \mbox{$\sum_{n=1}^{N}\mathrm{tr}(\mathbf{P}_{n}\mathbf{P}_{n}^{H})+\mathrm{tr}(\mathbf{F}_{n}\mathbf{F}_{n}^{H})$ $\leq \bar{P}_{t}$}, where \mbox{$\mathbf{P}_{n}=\left[\mathbf{p}_{c,n} \mathbf{p}_{1,n},\ldots,\mathbf{p}_{K,n}\right] $} and \mbox{$\mathbf{F}_{n}=\left[\mathbf{f}_{1,n},\ldots,\mathbf{f}_{L,n}\right] $}. 
The signal received by IU-$k$ on subcarrier-$n$ is 
\vspace{-0.2cm}
\begin{align}
	y_{k,n}&=\mathbf{h}_{k,n}^{H}\mathbf{x}_{n}+z_{k,n}, \quad  k \in \mathcal{K}, \  n\in \mathcal{S},
	\vspace{-0.3cm}
	\label{eqn:receive_signal}	
\end{align}
where \mbox{$\mathbf{h}_{k,n} \in \mathbb{C}^{n_{t}}$} is the channel vector of IU-$k$ on subcarrier-$n$ and \mbox{$z_{k,n} \sim \mathcal{CN}(0,1)$} is the Additive White Gaussian Noise (AWGN). Similarly, the signal received by AU-$l$ on subcarrier-$n$ is written as 
\vspace{-0.2cm}
\begin{align}
	r_{l,n}&=\mathbf{g}_{l,n}^{H}\mathbf{x}_{n}+\nu_{l,n}, \quad l \in \mathcal{L}, \  n\in \mathcal{S}, 
	\vspace{-0.1cm}
	\label{eqn:receive_signal_adversarial}	
\end{align}
where \mbox{$\mathbf{g}_{l,n} \in \mathbb{C}^{n_{t}}$} is the channel vector of AU-$l$ on subcarrier-$n$ and \mbox{$\nu_{l,n} \sim \mathcal{CN}(0,1)$} is the AWGN.

At the receiver side, detection of the messages is carried out using Successive Interference Cancellation (SIC). The common stream is detected first to obtain the common message estimate $\hat{W}_{c,n}$ by treating the private streams as noise. The common stream is then reconstructed using $\hat{W}_{c,n}$ and subtracted from the received signal. The remaining signal is used to detect the private messages $\hat{W}_{p,k,n}$. Finally, the estimated message for IU-$k$, $\hat{W}_{k,n}$, is obtained by combining $\hat{W}_{c,k,n}$ and $\hat{W}_{p,k,n}$. 
We write the Signal-to-Interference-plus-Noise Ratio (SINR) expressions for the common and private streams at IU-$k$ as
\begin{align}
	\gamma_{c,k,n}\hspace{-0.1cm}=\hspace{-0.1cm}\frac{|\mathbf{h}_{k,n}^{H}\mathbf{p}_{c,n}|^{2}}{1+Z_{c,k,n}+J_{k,n}}, \quad 
	\gamma_{k,n}\hspace{-0.1cm}=\hspace{-0.1cm}\frac{|\mathbf{h}_{k,n}^{H}\mathbf{p}_{k,n}|^{2}}{1+Z_{k,n}+J_{k,n}}, \nonumber
\end{align} 
\hspace{-0.05cm}with $Z_{c,k,n}=\sum_{i \in \mathcal{K}}|\mathbf{h}_{k,n}^{H}\mathbf{p}_{i,n}|^{2}$, $J_{k}=\sum_{j \in \mathcal{L}}|\mathbf{h}_{k,n}^{H}\mathbf{f}_{j,n}|^{2}$ and $Z_{k,n}=\sum_{i \in \mathcal{K}, i \neq k}|\mathbf{h}_{k,n}^{H}\mathbf{p}_{i,n}|^{2}$.

In this work, we consider the notion of jamming in the context of denial of service for the AUs. Our aim is to efficiently focus power on the AUs to disrupt the correct detection and decoding of their intended data transmissions from other users in their corresponding network. Our performance criterion is the focused power on pilot subcarrier-$n$ of an AU-$l$, $n \in \mathcal{S}_{p}$ and $l \in\mathcal{L}$, expressed as 
\begin{align}
	\Lambda_{l,n}=|\mathbf{g}_{l,n}^{H}\mathbf{p}_{c,n}|^{2}+\sum_{k \in \mathcal{K}}|\mathbf{g}_{l,n}^{H}\mathbf{p}_{k,n}|^{2}+\sum_{l^{\prime} \in \mathcal{L}}|\mathbf{g}_{l,n}^{H}\mathbf{f}_{l^{\prime},n}|^{2}.\nonumber
\end{align}
We assume that the transmitter has synchronisation with the AU transmissions \cite{karlsson_2014, karlsson_2017, miller_2012} and a perfect knowledge of $\mathcal{S}_{p}$ \cite{Shahriar_2015, clancy_2011, miller_2012, patel_2004, han_2008}.
 
We consider the practical case where the transmitter does not have access to perfect Channel State Information (CSI). The channel model of IU-$k$ is expressed as
\begin{align}
	\mathbf{h}_{k,n}=\sqrt{1\minus\sigma_{ie}^{2}}\widehat{\mathbf{h}}_{k,n}+\sigma_{ie}\widetilde{\mathbf{h}}_{k,n},
\end{align}
where $\widehat{\mathbf{h}}_{k,n}$ is the estimate of the channel on subcarrier-$n$ at the transmitter and $\widetilde{\mathbf{h}}_{k,n}$ is the channel estimation error with i.i.d. complex Gaussian elements of unit variance. The covariance matrix of the channel of AU-$l$ on subcarrier-$n$ is expressed as
 $\mathbf{R}_{g_{l,n}}=\mathbb{E}\left\lbrace\mathbf{g}_{l,n}\mathbf{g}_{l,n}^{H}\right\rbrace$.
We assume that the channel is fixed during the transmission of an MC waveform block. We also assume perfect CSI at the receivers.

\section{RSMA for Joint Communications and Jamming with OFDM Waveform}
In this section, we give the problem formulation to obtain the optimal precoders for the system model in Section~\ref{sec:system}. 
Our objective is to maximize the ergodic mutual information under imperfect CSIT for IUs while focusing a certain amount of jamming power on the pilot subcarriers of the AUs. 
The receiver employs carrier non-cooperative processing of the MC waveform. Such approach considers an independent processing of each subcarrier at the receiver, which is suitable for practical scenarios due to its low complexity \cite{Palomar_2003}. 

We define the matrices $\mathbf{H}_{k}=\mathrm{Diag}( \mathbf{h}_{k,1},  \ldots, \mathbf{h}_{k,N})$, $\mathbf{P}=\mathrm{Diag}( \mathbf{P}_{1},  \ldots, \mathbf{P}_{N})$ and $\mathbf{Z}_{k}=\mathrm{Diag}((Z_{k,1}\hspace{-0.1cm}+\hspace{-0.1cm}J_{k,1}\hspace{-0.1cm}+\hspace{-0.1cm}N_{0}),\ldots,(Z_{k,N}\hspace{-0.1cm}+\hspace{-0.1cm}J_{k,N}\hspace{-0.1cm}+\hspace{-0.1cm}N_{0}) )$. Under the assumption of carrier non-cooperative processing, the mutual information at IU-$k$ is expressed as $I_{k}=\log|\mathbf{I}+\mathbf{Z}_{k}^{-1}\mathbf{H}^{H}_{k}\mathbf{P}\mathbf{P}^{H}\mathbf{H}_{k}|$ \cite{Palomar_2003, Cover_1991}. 

In order to obtain the optimal precoders that maximize the mutual information, we make use of the mutual information - Mean Square Error (MSE) relations. We note that in addition to the numerous works, such approach is taken for designing the optimal precoders for MC multi-antenna systems in \cite{Palomar_2003} and RSMA in MISO BC in \cite{Joudeh_2016}. We first obtain the optimal receive filter, $g_{k,n}$, that minimizes the Mean Square Error (MSE) \mbox{$\mathbb{E}\left\lbrace|\epsilon_{k,n}|^{2}\right\rbrace=\mathbb{E}\left\lbrace|g_{k,n}y_{k,n}-x_{k,n}|^{2} \right\rbrace$}, $\forall k \in \mathcal{K}$, $\forall n \in \mathcal{S}$. It is well known that the solution is given by a Minimum MSE (MMSE) filter 
\begin{align}
	g^{\mathrm{opt}}_{k,n}=\mathbf{p}^{H}_{k,n}\mathbf{h}_{k,n}\left( |\mathbf{h}_{k,n}^{H}\mathbf{p}_{k,n}|^{2}+Z_{k,n}+J_{k,n}+N_{0}\right) ^{-1}.\nonumber
\end{align}
The resulting MSE is expressed as
\begin{align}
	\epsilon^{\mathrm{opt}}_{k,n}\hspace{-0.1cm}=\hspace{-0.1cm}\left( |\mathbf{h}_{k,n}^{H}\mathbf{p}_{k,n}|^{2}\hspace{-0.1cm}+\hspace{-0.1cm}Z_{k,n}\hspace{-0.1cm}+\hspace{-0.1cm}J_{k,n}\hspace{-0.1cm}+\hspace{-0.1cm}N_{0}\right) ^{-1}\hspace{-0.1cm}(Z_{k,n}\hspace{-0.1cm}+\hspace{-0.1cm}J_{k,n}\hspace{-0.1cm}+\hspace{-0.1cm}N_{0}).
	\label{eqn:mse}
\end{align}
The mutual information-MSE relation is given by \mbox{$I_{k}=-\log|\mathbf{E}_{k}|$}, where $\mathbf{E}_{k}=\mathrm{Diag}(\epsilon^{\mathrm{opt}}_{k,1}, \ldots, \epsilon^{\mathrm{opt}}_{k,N} )$ \cite{Palomar_2003}, which can be expanded as
\begin{align}
	\hspace{-0.1cm}I_{k}\hspace{-0.1cm}=\hspace{-0.1cm}-\log\left( \prod_{n=1}^{N}\epsilon^{\mathrm{opt}}_{k,n}\right)\hspace{-0.1cm}=\hspace{-0.1cm}-\sum_{n=1}^{N}\log(\epsilon^{\mathrm{opt}}_{k,n})\hspace{-0.1cm}=\hspace{-0.1cm}\sum_{n=1}^{N}I_{k,n}. \label{eqn:mutual_private}
\end{align}

In the context of RSMA, \eqref{eqn:mutual_private} represents the mutual information for the private stream of IU-$k$ and has the requirement of being decodable by the corresponding user. 
On the other hand, the common stream has a stricter requirement of being decodable by all IUs in the system. 
In order to satisfy such a requirement, we consider the mutual information per subcarrier for the common stream, so that the decodability of the message on each subcarrier is guaranteed by all IUs\footnote{By carrier cooperative processing, the decodability can be guaranteed over $I_{c,k}$ instead of $I_{c,k,n}$ with proper coding methods \cite{Raleigh_1998}.}. We define $I_{c,k,n}=-\log(\epsilon^{\mathrm{opt}}_{c,k,n})$,
where $\epsilon^{\mathrm{opt}}_{c,k,n}$ is obtained by replacing $\mathbf{p}_{k,n}$ and $Z_{k,n}$ in \eqref{eqn:mse} by $\mathbf{p}_{c,n}$ and $Z_{c,k,n}$, respectively. 

Next, we determine the jamming power constraint for statistical CSIT. We consider the average power focused on subcarrier-$n$ of AU-$l$ \cite{Xing_2013}. Accordingly, 
\vspace{-0.3cm}
\begin{align}
	\mathbb{E}\hspace{-0.05cm}&\left\lbrace\hspace{-0.05cm}|\mathbf{g}_{l,n}^{H}\mathbf{p}_{c,n}|^{2}\hspace{-0.1cm}+\hspace{-0.1cm}\sum_{k \in \mathcal{K}}\hspace{-0.1cm}|\mathbf{g}_{l,n}^{H}\mathbf{p}_{k,n}|^{2}\hspace{-0.1cm}+\hspace{-0.1cm}\sum_{l^{\prime} \in \mathcal{L}}\hspace{-0.1cm}|\mathbf{g}_{l,n}^{H}\mathbf{f}_{l^{\prime},n}|^{2}\hspace{-0.05cm}\right\rbrace  \nonumber \\
	&=\mathbf{p}_{c,n}^{H}\mathbf{R}_{g_{l,n}}\mathbf{p}_{c,n}\hspace{-0.15cm}+\hspace{-0.15cm}\sum_{k \in \mathcal{K}}\mathbf{p}_{k,n}^{H}\mathbf{R}_{g_{l,n}}\mathbf{p}_{k,n}
	\hspace{-0.15cm}+\hspace{-0.15cm}\sum_{l^{\prime} \in \mathcal{L}}\mathbf{f}_{l^{\prime},n}^{H}\mathbf{R}_{g_{l,n}}\mathbf{f}_{l^{\prime},n} \nonumber \\
	&\triangleq \bar{\Lambda}_{l,n}(\mathbf{P}_{n},\mathbf{F}_{n}).
	\vspace*{-0.5cm}
	\label{eqn:average_power}
\end{align}

\vspace{-0.2cm}
We formulate the optimization problem as
\vspace{-0.2cm}
\begin{subequations}
	\begin{alignat}{3}
		\hspace{-0.4cm}\max_{\mathbf{P},\mathbf{F},\mathbf{c}}&     \quad  \sum_{n \in \mathcal{S}}\sum_{k \in \mathcal{K}}\left(C_{k,n}+I_{k,n}(\mathbf{P}_{n},\mathbf{F}_{n})\right)      \label{eqn:obj}   \\
		\text{s.t.}&  \quad  \sum_{k \in \mathcal{K}}C_{k,n} \leq I_{c,k,n}(\mathbf{P_{n}},\mathbf{F_{n}}), \forall n \in \mathcal{S},  \forall k \in \mathcal{K} \label{eqn:common_rate_final} \\
		&\quad  \bar{\Lambda}_{l,n}(\mathbf{P}_{n},\mathbf{F}_{n}) \geq J^{thr}_{l,n},  \quad \forall n \in \mathcal{S}_{p}, \ \forall l \in \mathcal{L} \label{eqn:energy_3} \\
		& \quad  \sum_{n \in \mathcal{S}} \mathrm{tr}(\mathbf{P}_{n}\mathbf{P}_{n}^ {H}) + \mathrm{tr}(\mathbf{F}_{n}\mathbf{F}_{n}^ {H}) \leq \bar{P}_{t},  \label{eqn:total_power} \\
		& \quad   \mathbf{c} \geq \mathbf{0} \label{eqn:c}.
	\end{alignat}
	\label{eqn:problem1}
\end{subequations}
\hspace{-0.25cm}where $\mathbf{c}=\left[C_{1,1},\ldots,C_{K,1},\ldots,C_{1,N},\ldots,C_{K,N}\right]^{T} $ with $C_{k,n}$ being the portion of the common mutual information for IU-$k$ on subcarrier-$n$.  
Rate-MSE transformations similar to the one in \eqref{eqn:mutual_private} have been used in \cite{Joudeh_2016} without an MC waveform to transform the non-convex sum-rate maximization problem for RSMA into a convex one. Therefore, we follow the approach in \cite{Joudeh_2016} and define the augmented MSEs \mbox{$\xi_{c,k,n}=u_{c,k,n}\epsilon_{c,k,n}-\log_{2}(u_{c,k,n})$} and \mbox{$\xi_{k,n}=u_{k,n}\epsilon_{k,n}-\log_{2}(u_{k,n})$} to obtain $\xi^{\mathrm{opt}}_{k,n}=1\minus I_{k,n}$ and $\xi^{\mathrm{opt}}_{c,k,n}=1\minus I_{c,k,n}$, $\forall n \in \mathcal{S}$.
The resulting formulation is
\begin{subequations}
			\vspace{-0.3cm}	
			\begin{alignat}{3}
		\min_{\mathbf{P},\mathbf{F},\mathbf{x},\mathbf{u},\mathbf{g}}&     \quad  \sum_{n \in \mathcal{S}}\sum_{k \in \mathcal{K}}\left(X_{k,n}+ \xi_{k,n}(\mathbf{P}_{n},\mathbf{F}_{n})\right)      \label{eqn:obj_f}   \\
		\text{s.t.}&  \quad  \sum_{k \in \mathcal{K}}\hspace{-0.08cm}X_{k,n} \hspace{-0.08cm}+\hspace{-0.08cm} 1 \geq \xi_{c,k,n}(\mathbf{P}_{n},\mathbf{F}_{n}), \nonumber \\ 
		&\quad\quad\quad\quad\quad\quad\quad\quad\quad\quad\ \forall n \in \mathcal{S}, \forall k \in \mathcal{K}  \label{eqn:common_rate_1_f} \\
		&\quad  \bar{\Lambda}_{l,n}(\mathbf{P}_{n},\mathbf{F}_{n})  \geq J^{thr}_{l,n}, \forall n \in \mathcal{S}_{p}, \ \forall l \in \mathcal{L} \label{eqn:energy_3_f} \\
		& \quad  \sum_{n \in \mathcal{S}} \mathrm{tr}(\mathbf{P}_{n}\mathbf{P}_{n}^ {H}) + \mathrm{tr}(\mathbf{F}_{n}\mathbf{F}_{n}^ {H}) \leq \bar{P}_{t},  \label{eqn:total_power_f} \\
		& \quad   \mathbf{x} \leq \mathbf{0}, \label{eqn:x}
	\end{alignat}
	\label{eqn:problem1_final}
\end{subequations}
\hspace{-0.15cm}where \mbox{$\mathbf{x}=\left[X_{1,1},\ldots,X_{K,1}, \ldots, X_{1,N},\ldots,X_{K,N}\right]^{T} $} and $X_{k,n}=-C_{k,n}$. For the sake of brevity, we skip detailed derivations to extend the problem formulation for the imperfect CSIT case by the Sample Average Approximation (SAA) and refer the interested reader to \cite{Joudeh_2016}. 

We note that the constraint \eqref{eqn:energy_3_f} is not convex due to the convex function $\bar{\Lambda}_{l,n}(\mathbf{P}_{n})$  constrained to a lower bound. We follow the approach followed in \cite{Mao_2016} and obtain a convex constraint by using first-order Taylor expansion for the quadratic function in \eqref{eqn:energy_3_f}. Specifically, one can lower bound the term $\mathbf{p}_{k,n}^{H}\mathbf{R}_{g_{l,n}}\mathbf{p}_{k,n}$ at a point $\mathbf{p}^{t}_{k,n}$ as
\vspace{-0.2cm}
\begin{align}
	\mathbf{p}_{k,n}^{H}&\mathbf{R}_{g_{l,n}}\mathbf{p}_{k,n}	 \nonumber\\
	&\geq 2\mathrm{Re}\left\lbrace(\mathbf{p}_{k,n}^{t})^{H}\mathbf{R}_{g_{l,n}}\mathbf{p}_{k,n} \right\rbrace-(\mathbf{p}_{k,n}^{t})^{H}\mathbf{R}_{g_{l,n}}\mathbf{p}^{t}_{k,n} \nonumber\\
	&\triangleq \bar{\phi}^{t}(\mathbf{p}_{k,n},\mathbf{R}_{l,n})
	\label{eqn:avg_energy_convex}
\end{align}
Using the function in \eqref{eqn:avg_energy_convex}, we write
\begin{align}
	\bar{\Lambda}_{l,n}&(\mathbf{P}_{n},\mathbf{F}_{n}) \geq  \nonumber \\
	& \bar{\phi}^{t}(\mathbf{p}_{c,n},\mathbf{R}_{l,n})\hspace{-0.08cm}+\hspace{-0.1cm}\sum_{k \in \mathcal{K}}\hspace{-0.08cm}\bar{\phi}^{t}(\mathbf{p}_{k,n},\mathbf{R}_{l,n})\hspace{-0.08cm}+\hspace{-0.1cm}\sum_{l \in \mathcal{L}}\hspace{-0.08cm}\bar{\phi}^{t}(\mathbf{f}_{l,n},\mathbf{R}_{l,n}) \nonumber \\
	&\triangleq\bar{\Lambda}^{t}_{l,n}(\mathbf{P}_{n},\mathbf{F}_{n}) \nonumber
\end{align}
\begin{figure}[t!]
	\removelatexerror
	\begin{algorithm}[H]
		\caption{SCA-based algorithm}
		\label{alg:sca}
		$t \gets 0$, $i \gets 0$, $\mathrm{WSR}^{0} \gets 0$, $\hat{\mathbf{P}}$, $\hat{\mathbf{F}}$   \\
		\While{$|\mathrm{WSR}^{i}-\mathrm{WSR}^{i-1}|> \epsilon_{r}$}{
			$\boldsymbol{u}^{i} \gets$ updateWeights($\hat{\mathbf{P}}$, $\hat{\mathbf{F}}$) \\
			$\boldsymbol{g}^{i} \gets$ updateFilters($\hat{\mathbf{P}}$, $\hat{\mathbf{F}}$) \\
			\While{$|\mathrm{WMMSE}^{t}-\mathrm{WMMSE}^{t-1}|> \epsilon_{m}$}{
				($\mathbf{P}^{t+1}$, $\mathbf{F}^{t+1}$, $\mathbf{x}^{t+1}$) $\gets$   \eqref{eqn:problem1_final_convex} for given  $\boldsymbol{u}^{i}$, $\boldsymbol{g}^{i}$ and ($\mathbf{P}^{t}$, $\mathbf{F}^{t}$)  \\
				$\mathrm{WMMSE}^{t+1} \gets$ updateWMMSE($\mathbf{P}^{t+1}$, $\mathbf{F}^{t+1}$) \\
				$t \gets t + 1$\\
			}
			$\hat{\mathbf{P}} \gets \mathbf{P}^{t}$,\ $\hat{\mathbf{F}} \gets \mathbf{F}^{t}$ \\
			$\mathrm{WSR}^{i+1} \gets$ updateWSR($\hat{\mathbf{P}}$, $\hat{\mathbf{F}}$) \\
			$t \gets 0$, \ $i \gets i + 1$\\
		}
		\Return ($\hat{\mathbf{P}}$, $\hat{\mathbf{F}}$)
	\end{algorithm}
	\vspace{-0.7cm}
\end{figure}
The final problem formulation is written as
\begin{subequations}
	\begin{alignat}{3}
		\min_{\mathbf{P},\mathbf{F},\mathbf{x},\mathbf{u},\mathbf{g}}&     \quad  \sum_{n \in \mathcal{S}}\sum_{k \in \mathcal{K}}\left(X_{k,n}+ \xi_{k,n}(\mathbf{P}_{n},\mathbf{F}_{n})\right)      \label{eqn:obj_f_convex}   \\
		\text{s.t.}& \quad   \bar{\Lambda}^{t}_{l,n}(\mathbf{P}_{n}, \mathbf{F}_{n}) \geq J^{thr}_{l,n},   n \in \mathcal{S}_{p}, \ l \in \mathcal{L} \label{eqn:energy_4_f} \\
		&\quad \eqref{eqn:common_rate_1_f},\ \eqref{eqn:total_power_f} \nonumber \ \eqref{eqn:x}.
	\end{alignat}
	\label{eqn:problem1_final_convex}
\end{subequations}
\hspace{-0.1cm}The SCA-based algorithm to solve \eqref{eqn:problem1_final_convex} is given in Alg.~\ref{alg:sca}. 

We set the power threshold on subcarrier-$n$ of AU-$l$ as
\begin{align}
	J^{thr}_{l,n}=\rho \frac{\bar{P}_{t}}{|S_{p}|L}\tau_{l,n},  \quad \forall n \in \mathcal{S}_{p}L, \ \forall l \in \mathcal{L},
\end{align}
where $\rho \in \left[0,1 \right]$ determines the strictness of the jamming power threshold by adjusting it to a percentage of the maximum value, \mbox{$|S_{p}|$} denotes the cardinality of the set $S_{p}$ and $\tau_{n,l}$ is the largest eigenvalue of the matrix $\mathbf{R}_{g_{l,n}}$.

\section{Simulation Results}
\label{sec:simulation}
We perform simulations to demonstrate the sum-rate performance achieved by RSMA and SDMA using the optimized precoders. Note that the optimal precoders for SDMA can be obtained by turning off the common stream in the optimization problem formulation. We set $n_{t}=4$, $K=2$ and $L=1$ for the system parameters. We use Cyclic-Prefix (CP)-OFDM waveform with $32$ subcarriers and a CP length of $10\mu$s. The error variance of the channels of the IUs is modeled as $\sigma_{ie}^{2}=(\bar{P}_{t}/N)^{-\alpha}$ , where $\alpha$ is the CSIT quality scaling factor \cite{Yang_2013, Joudeh_2016}, and is set as $\alpha=0.6$ in the simulations. 
We define the private rate of IU-$k$ for an MC waveform as \mbox{$R_{k}=I_{k}/N$}, $\forall k\in \mathcal{K}$, following the formulation in \cite{Raleigh_1998} for the carrier cooperative case. This serves as an upper bound for the non-cooperative case since carrier cooperative processing is a more general model \cite{Palomar_2003}. Accordingly, the common rate is defined as \mbox{$R_{c,k}=\frac{1}{N}\sum_{k \in \mathcal{K}}C_{k}$}. The sum-rate for RSMA is calculated as \mbox{$R_{\mathrm{sum}}=\sum_{k \in \mathcal{K}}\left( \frac{C_{k}}{N}+R_{k}\right) $}.

\begin{figure}[t!]
	\centering
	\begin{subfigure}[t]{.25\textwidth}
		\centerline{\includegraphics[width=1.8in,height=1.8in,keepaspectratio]{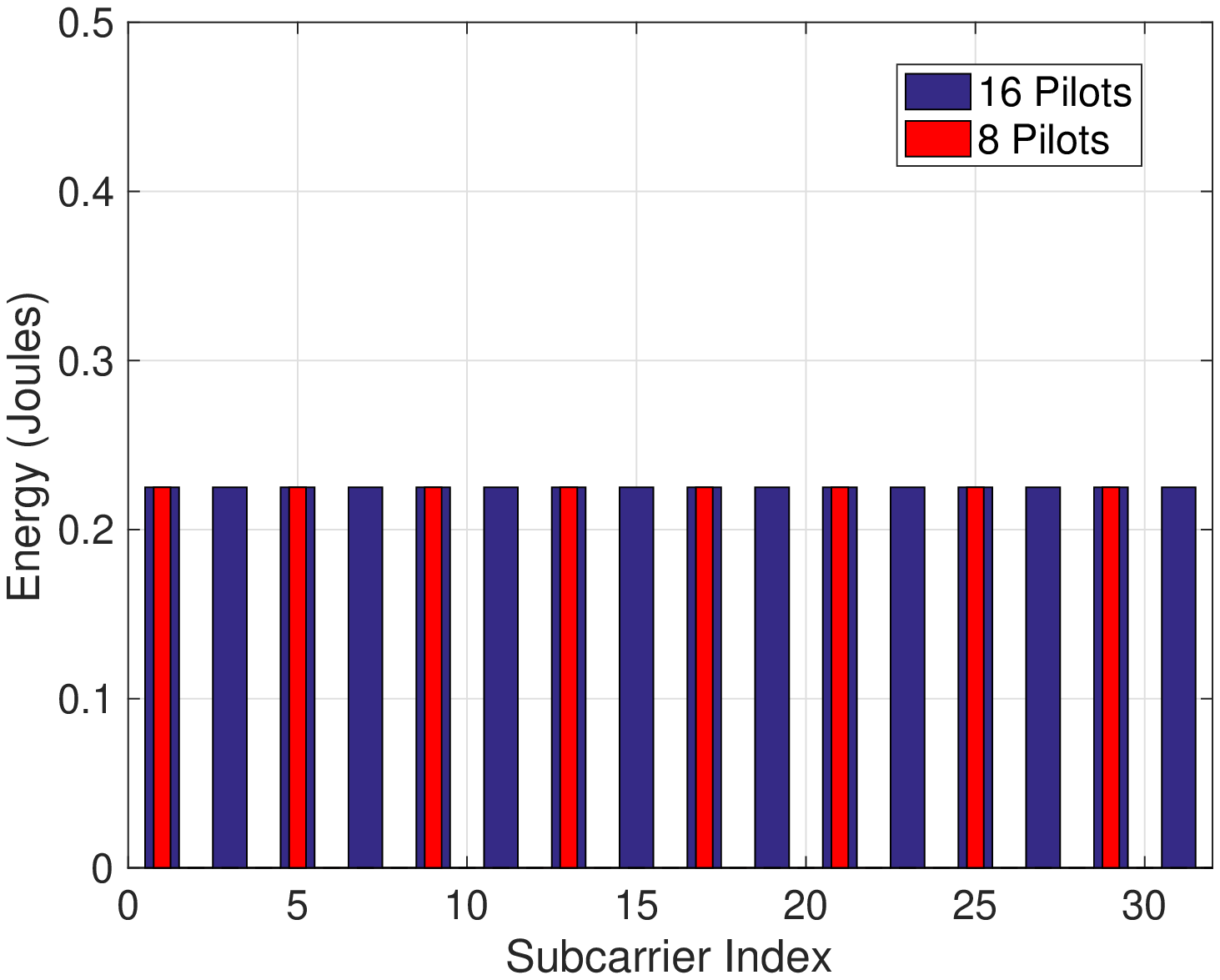}}
		\caption{Strategy 1  \mbox{($\rho=0.9\frac{|\mathcal{S}_{p}|}{N/2}$)}.}
		\label{fig:Threshold_Cnst}
	\end{subfigure}%
	\begin{subfigure}[t]{.25\textwidth}
		\centerline{\includegraphics[width=1.8in,height=1.32in]{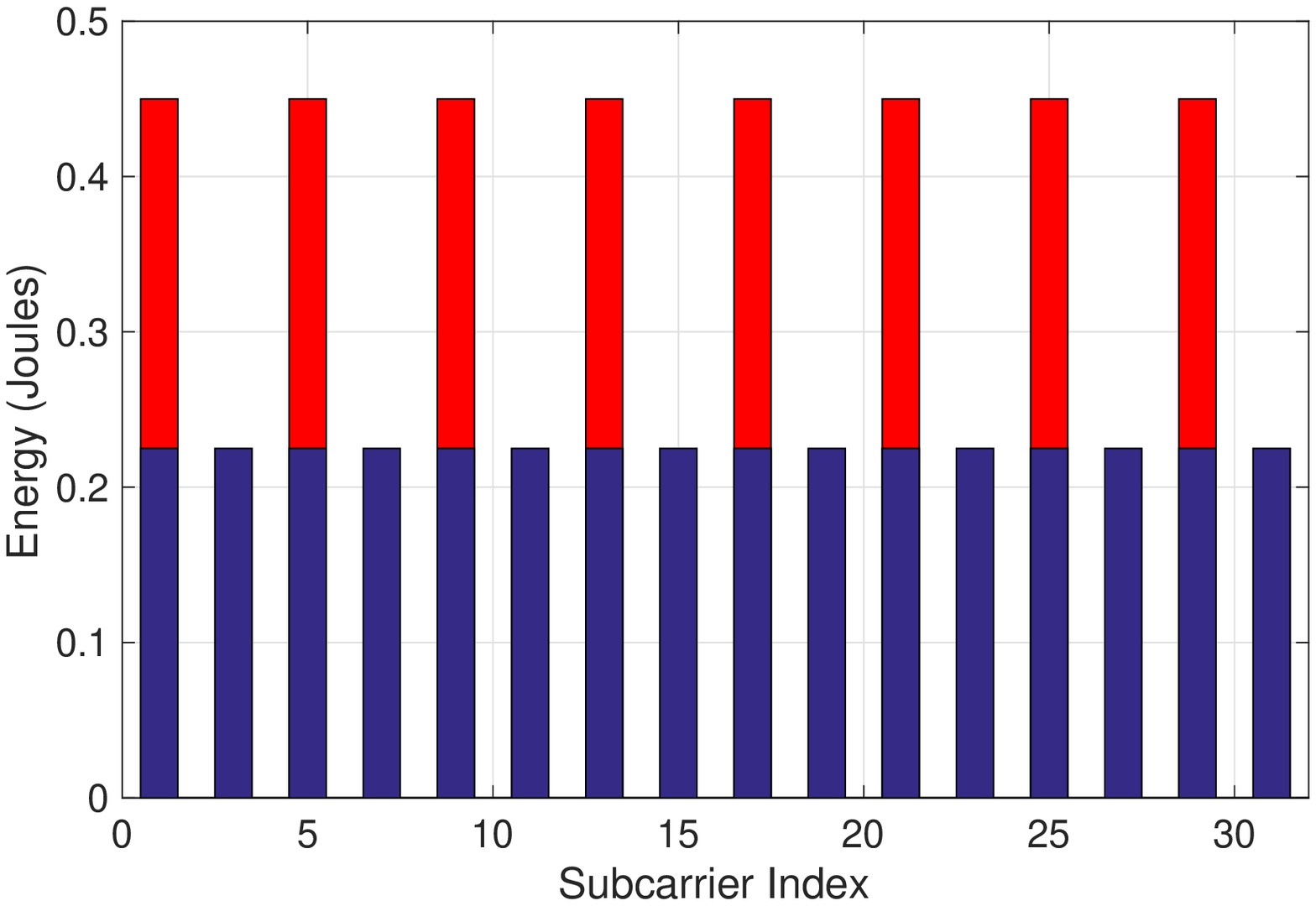}}
		\caption{Strategy 2 \mbox{($\rho=0.9$)}.}
		\label{fig:Threshold_Vary}
	\end{subfigure}
	\vspace{-0.2cm}
	\caption{Power thresholds with different strategies.}
	\label{fig:thresholds}
	\vspace{-0.5cm}
\end{figure}

We consider two different strategies to set the threshold for the jamming power. Fig.~\ref{fig:thresholds} illustrates the jamming threshold values $J^{thr}_{1,n}$ obtained by the proposed strategies. In Strategy 1, the threshold coefficient $\rho$ is varied proportional to the number of pilots, such that $\rho \propto |\mathcal{S}_{p}|$. 
Such method ensures that the total jamming power focused on an AU varies with the number of pilots, while the focused jamming power threshold per subcarrier stays the same. In Strategy 2, the threshold coefficient $\rho$ is kept constant irrespective of the number of pilots. In this case, the focused jamming power threshold varies with varying number of pilots, which leads to increased focused jamming power per subcarrier when the number of pilots decreases (or vice versa).

We start our analysis with basic multi-antenna channel models in the form of \mbox{$\mathbf{h}_{1,n}=[1,1,1,1]^{H}$}, \mbox{$\mathbf{h}_{2,n}=[1,e^{j\theta},e^{j2\theta},e^{j3\theta}]^{H}$} with $\theta=4\pi/9$ and \mbox{$\mathbf{g}_{1,n}=[1,e^{j\beta},e^{j2\beta},e^{j3\beta}]^{H}$} with $\beta=2\pi/9$, $\forall n \in \mathcal{S}$. Note that the channel is not frequency-selective in the considered model. {i.e.,} the channel frequency response is identical on each subarrier.
The second-order statistics of the channel $\mathbf{g}_{1,n}$ is obtained by averaging the correlation matrix over $\beta \in \left[0,+4\pi/9\right] $. 
Fig.~\ref{fig:thresholds} shows the jamming power thresholds for the two strategies described above.

\begin{figure}[t!]
	\centering
	\begin{subfigure}[t]{.25\textwidth}
		\centerline{\includegraphics[width=1.8in,height=1.8in,keepaspectratio]{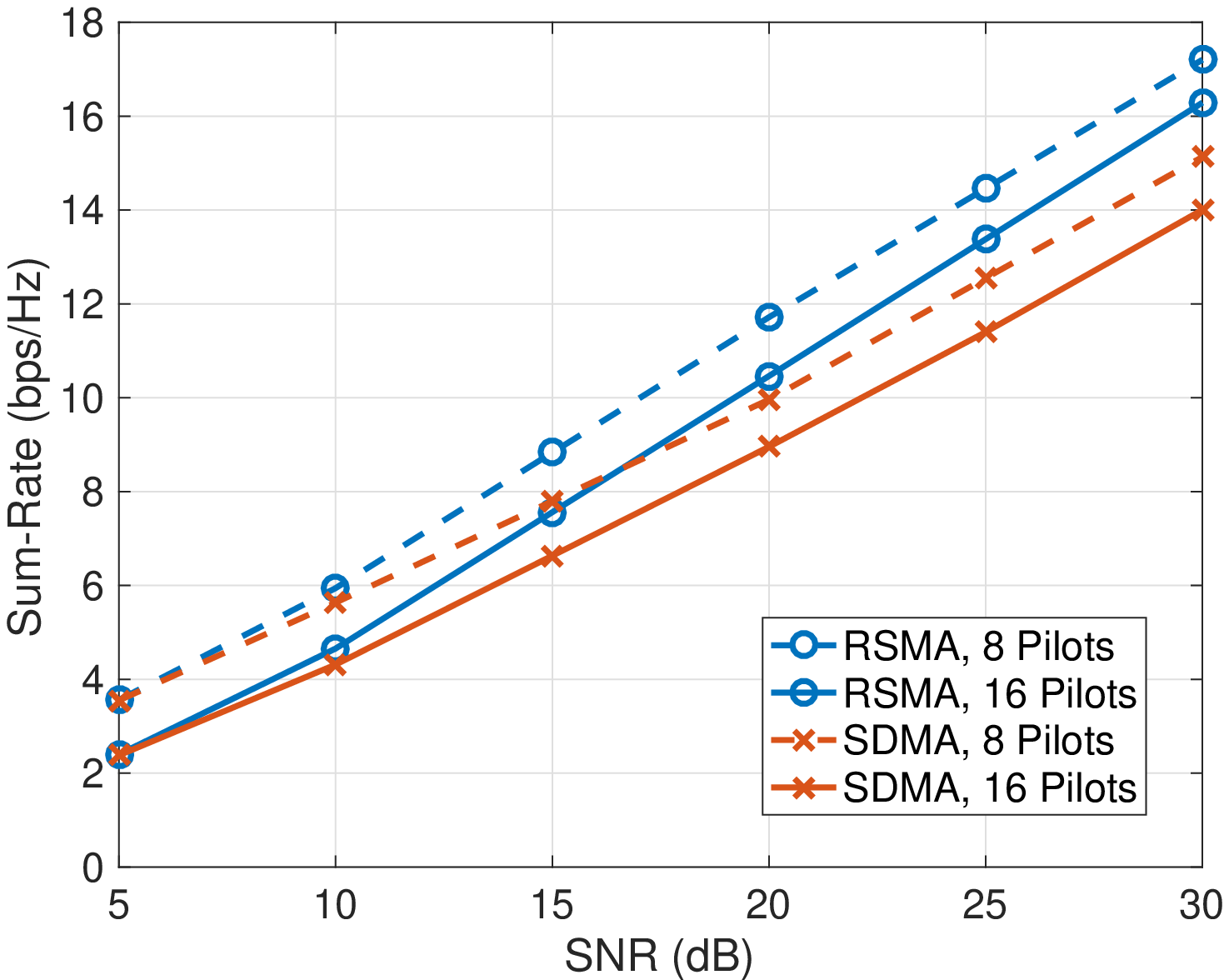}}
		\caption{Sum-rate vs. SNR, strategy 1.}
		\label{fig:SNRvsSR_ThrCnst}
	\end{subfigure}%
	\begin{subfigure}[t]{.25\textwidth}
		\centerline{\includegraphics[width=1.8in,height=1.8in,keepaspectratio]{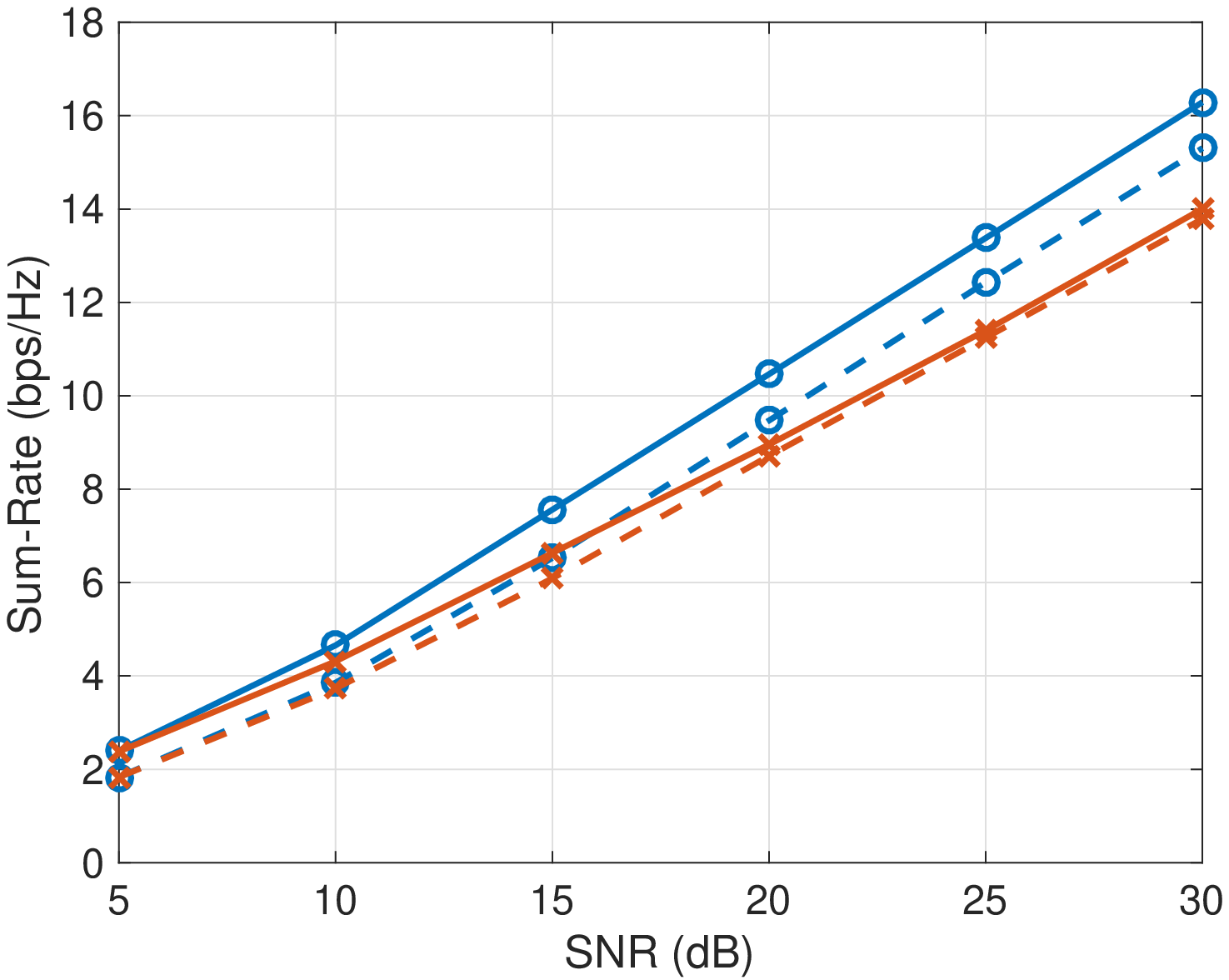}}
		\caption{Sum-rate vs. SNR, strategy 2.}
		\label{fig:SNRvsSR_ThrVary}
	\end{subfigure}
	\vspace{-0.2cm}
	\caption{Sum-rate vs. SNR with different strategies.}
	\vspace{-0.7cm}
\end{figure}
Fig.~\ref{fig:SNRvsSR_ThrCnst} shows the sum-rate performance of RSMA and SDMA for jamming power threshold according to Strategy 1 as depicted in Fig.~\ref{fig:Threshold_Cnst}. 
It is observed from the figure that the sum-rate decreases with increasing number of pilot subcarriers to jam as expected.
The decrease in the sum-rate performance of SDMA is higher than that of RSMA, implying that RSMA achieves higher sum-rate on the subcarriers used for joint communications and jamming. 

Fig.~\ref{fig:SNRvsSR_ThrVary} shows the sum-rate performance of RSMA and SDMA with respect to SNR for jamming power threshold set according to Strategy 2 as depicted in Fig.~\ref{fig:Threshold_Vary}. 
In contrast to the case in Fig.~\ref{fig:SNRvsSR_ThrCnst} the sum-rate increases as the number of pilots increases for both RSMA and SDMA. Such behaviour is explained as follows.
Recall that the sum-rate is defined as the average of mutual information over all subcarriers. In jamming strategy 2, the portion of the total transmit power allocated to the pilot subcarriers does not change significantly with varying number of pilots. Similarly, the remaining transmit power distributed among the data subcarriers is approximately constant with varying number of data subcarriers. Consequently, the mutual information over data subcarriers decrease with increasing number of data subcarriers . Therefore, the use case with $8$ pilot and $24$ data subcarriers achieve less sum-rate than the one with $16$ pilot and $16$ data subcarriers, when the sum-rate is calculated as the average mutual information over all subcarriers.
RSMA achieves a significantly higher sum-rate on the subcarriers with joint communications and jamming, thus resulting with an increase in the rate averaged over all subcarriers with increasing number of pilot subcarriers. On the other hand, SDMA achieves a lower rate on those subcarriers, thus the increase being much less significant compared to RSMA.  

\begin{figure}[t!]
	\centerline{\includegraphics[width=2.7in,height=2.7in,keepaspectratio]{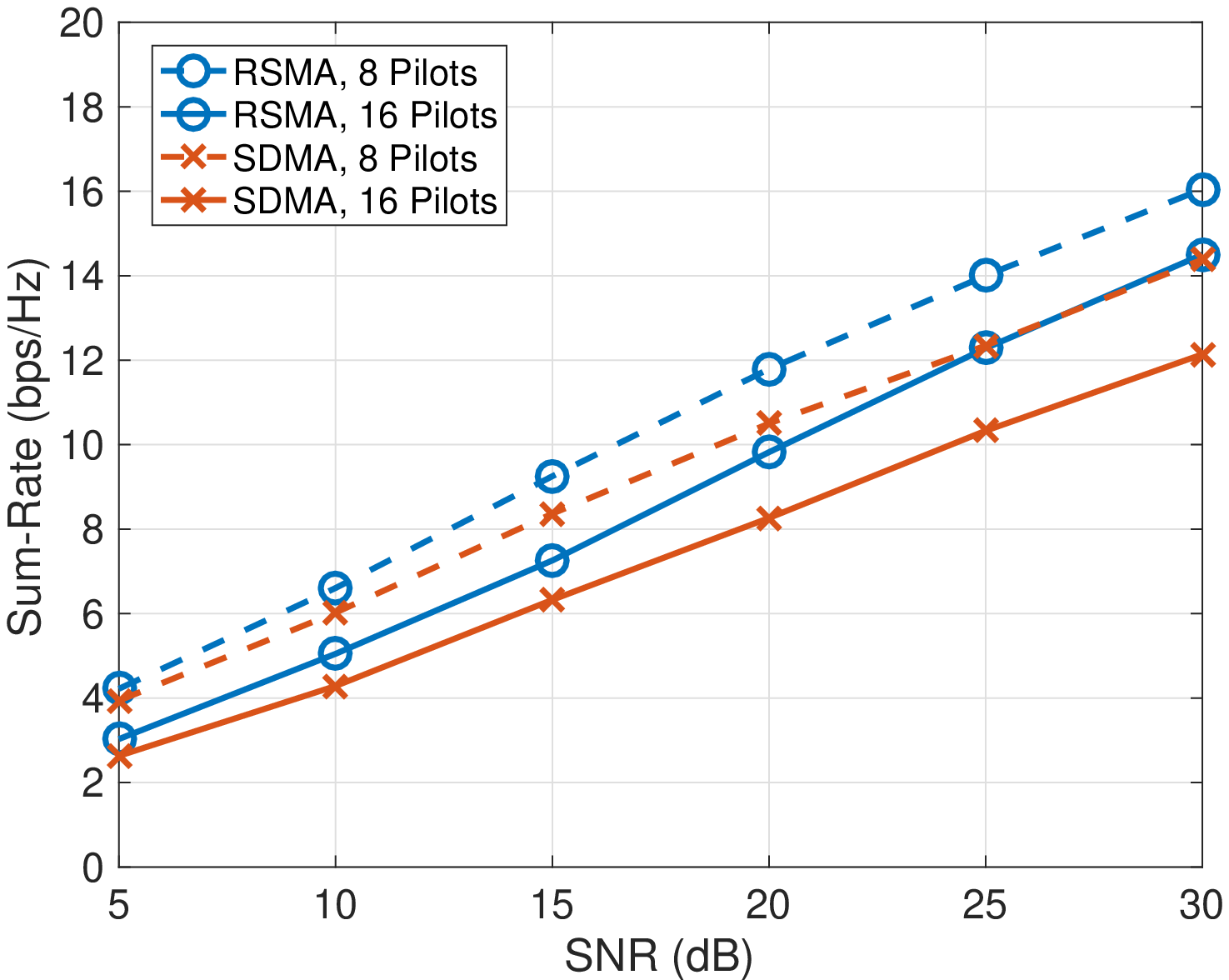}}
	\caption{Sum-rate vs. SNR, strategy 1, urban macro-cell.}
	\label{fig:Quadriga_ThrCnst}
	\vspace{-0.7cm}
\end{figure}
Next, we investigate the performance under a frequency-selective channel model. We use the Quadriga Channel Generator \cite{jaeckel_2019} to generate channels according to the 3GPP Urban Macro-Cell channel model. The channel each of IU have a delay spread of $1200$ns and $23$ clusters, with each cluster consisting of $20$ rays. The channel of the AU have a delay spread of $1200$ns with identical numbers of clusters and rays. The OFDM subcarrier spacing is set as $60$kHz to observe the frequency-selectivity in the considered channel model. Fig.~\ref{fig:Quadriga_ThrCnst} shows the sum-rate performance of RSMA and SDMA under such channel model. The results show that RSMA has improved sum-rate performance with respect to SDMA under realistic frequency-selective channels. 




\section{Conclusion}
We design optimal precoders for joint communications and jamming using RSMA with an MC waveform. We formulate the optimization problem as the maximization of the mutual information with a minimum jamming power constraint on pilot subcarriers of the AUs while performing data transmission to IUs. We consider the practical case of imperfect CSIT for the IUs and statistical CSIT for the AUs. By simulation results, we show that RSMA can achieve significantly higher sum-rate than SDMA while focusing the same amount of jamming power on AU pilot subcarriers. 



\vspace{-0.2cm}
\begin{thebibliography}{00}
\bibitem{clerckx_2016} B. Clerckx, H. Joudeh, C. Hao, M. Dai, and B. Rassouli, ``Rate splitting for MIMO wireless networks: A promising PHY-layer strategy for LTE evolution,'' \emph{IEEE Commun. Mag.}, vol. 54, no. 5, pp. 98–-105, May 2016.

\bibitem{clerckx_2019} B. Clerckx, Y. Mao, R. Schober and H. V. Poor, ``Rate-splitting unifying SDMA, OMA, NOMA, and multicasting in MISO broadcast channel: a simple two-user rate analysis,'' \emph{IEEE Wireless Commun. Lett.}, vol. 9, no. 3, pp. 349--353, Mar. 2020.

\bibitem{Joudeh_2016} H. Joudeh and B. Clerckx, ``Sum-rate maximization for linearly precoded downlink multiuser MISO systems with partial CSIT: a rate-splitting approach,'' \emph{IEEE Trans. Commun.}, vol. 64, no. 11, pp. 4847--4861, Nov. 2016.




\bibitem{mao_2018} Y. Mao, B. Clerckx and V. O. K. Li, ``Rate-splitting multiple access for downlink communication systems: bridging, generalizing, and outperforming SDMA and NOMA,'' \emph{EURASIP J. Wireless Commun. Netw.}, 2018.

\bibitem{mao_2019_2} Y. Mao, B. Clerckx, and V. O. K. Li, ``Rate-splitting for multi-antenna non-orthogonal unicast and multicast transmission: spectral and energy efficiency analysis,'' \emph{IEEE Trans. Commun.}, vol. 67, no. 12, pp. 8754–-8770, Dec 2019.

\bibitem{Yang_2013} S. Yang, M. Kobayashi, D. Gesbert, and X. Yi, ``Degrees of freedom of time correlated MISO broadcast channel with delayed CSIT,'' \emph{IEEE
Trans. Inf. Theory}, vol. 59, no. 1, pp. 315–328, Jan. 2013.

\bibitem{Shahriar_2015} C. Shahriar et al., ``PHY-Layer resiliency in OFDM communications: a tutorial,'' \emph{IEEE Commun. Surveys \& Tutorials}, vol. 17, no. 1, pp. 292-314, First quarter 2015.


\bibitem{clancy_2011} T. C. Clancy, ``Efficient OFDM denial: pilot jamming and pilot nulling,'' \emph{Proc. 2011 IEEE Int. Conf. Commun. (ICC)}, Kyoto, Japan, 2011, pp. 1--5.

\bibitem{miller_2012} R. Miller and W. Trappe, ``On the vulnerabilities of CSI in MIMO wireless communication systems,'' \emph{IEEE Trans. Mobile Comp.}, vol. 11, no. 8, pp. 1386--1398, Aug. 2012.

\bibitem{patel_2004} C. S. Patel, G. L. Stuber and T. G. Pratt, ``Analysis of OFDM/MC-CDMA under channel estimation and jamming,'' \emph{2004 IEEE Wireless Commun. Net. Conf.}, Atlanta, GA, USA, 2004, pp. 954--958, Vol.2.

\bibitem{han_2008} M. Han et al., ``OFDM channel estimation with jammed pilot detector under narrow-band jamming,'' \emph{IEEE Trans. Vehic. Tech.}, vol. 57, no. 3, pp. 1934--1939, May 2008.

\bibitem{karlsson_2014} M. Karlsson and E. G. Larsson, ``Massive MIMO as a cyber-weapon,'' \emph{2014 48th Asilomar Conf. Signals, Syst. Comp.}, Pacific Grove, CA, USA, 2014, pp. 661--665.

\bibitem{karlsson_2017} M. Karlsson, E. Björnson and E. G. Larsson, ``Jamming a TDD point-to-point link using reciprocity-based MIMO,'' \emph{Trans. Inf. Forensics Security}, vol. 12, no. 12, pp. 2957--2970, Dec. 2017.

\bibitem{Mao_2016} Y. Mao, B. Clerckx and V. O. K. Li, ``Rate-splitting for multi-user multi-antenna wireless information and power transfer,'' \emph{Proc. IEEE 20th Intern. Workshop Sign. Process. Adv. Wireless Commun. (SPAWC)}, Cannes, France, 2019, pp. 1--5.

\bibitem{Cover_1991} T. M. Cover and J. A. Thomas, \emph{Elements of Information Theory}. New York: Wiley, 1991.

\bibitem{Palomar_2003} D. P. Palomar, J. M. Cioffi and M. A. Lagunas, ``Joint tx-rx beamforming design for multicarrier MIMO channels: a unified framework for convex optimization,'' \emph{IEEE Trans. Signal Process.}, vol. 51, no. 9, pp. 2381--2401, Sept. 2003.

\bibitem{Xing_2013} C. Xing, N. Wang, J. Ni, Z. Fei and J. Kuang, ``MIMO beamforming designs with partial CSI under energy harvesting constraints,'' \emph{IEEE Signal Process. Lett.}, vol. 20, no. 4, pp. 363--366, April 2013.




\bibitem{Raleigh_1998} G. G. Raleigh and J. M. Cioffi, ``Spatio-temporal coding for wireless communication,'' \emph{IEEE Trans. Commun.}, vol. 46, no. 3, pp. 357--366, March 1998.





















\bibitem{jaeckel_2019} S. Jaeckel, L. Raschkowski, K. B\"{o}rner, L. Thiele, F. Burkhardt and E. Eberlein, ``QuaDRiGa - quasi deterministic radio channel generator, user manual and documentation,'' Fraunhofer Heinrich Hertz Institute, Tech. Rep. v2.2.0, 2019.


\end{thebibliography}
\end{document}